\documentclass[sn-mathphys-num]{sn-jnl}% Math and Physical Sciences Numbered Reference Style 
\usepackage{graphicx}%
\usepackage{multirow}%
\usepackage{amsmath,amssymb,amsfonts}%
\usepackage{amsthm}%
\usepackage{mathrsfs}%
\usepackage[title]{appendix}%
\usepackage{xcolor}%
\usepackage{textcomp}%
\usepackage{manyfoot}%
\usepackage{booktabs}%
\usepackage{algorithm}%
\usepackage{algorithmicx}%
\usepackage{algpseudocode}%
\usepackage{listings}%
\theoremstyle{thmstyleone}%
\theoremstyle{thmstyletwo}%

\theoremstyle{thmstylethree}%

\begin{document}

\title[Article Title]{Novel Experimental Platform to realize One-dimensional Quantum Fluids}


\author[1]{\fnm{Stephanie} \sur{McNamara}}\email{stmc4903@colorado.edu}
\author[2,3]{\fnm{Prabin} \sur{Parajuli}}\email{parparaj@iu.edu}
\author[3]{\fnm{Sutirtha} \sur{Paul}}\email{spaul19@vols.utk.edu}
\author[2]{\fnm{Garfield} \sur{Warren}}\email{gtwarren@iu.edu}
\author[3,4]{\fnm{Adrian} \sur{Del Maestro}}\email{adrian.delmaestro@utk.edu}
\author*[2]{\fnm{Paul E.} \sur{Sokol}}\email{pesokol@iu.edu}

\affil[1]{\orgdiv{College of Arts and Sciences}, \orgname{University of Colorado Boulder}, \orgaddress{\city{Boulder}, \postcode{80309}, \state{CO}, \country{USA}}}

\affil[2]{\orgdiv{Department of Physics}, \orgname{Indiana University}, \orgaddress{\city{Bloomington}, \postcode{47405}, \state{IN}, \country{USA}}}

\affil[3]{\orgdiv{Department of Physics}, \orgname{University of Tennessee}, \orgaddress{\city{Knoxville}, \postcode{37996}, \state{TN}, \country{USA}}}

\affil[4]{\orgdiv{Min Kao Department of Electrical Engineering and Computer Science}, \orgname{University of Tennessee}, \orgaddress{\city{Knoxville}, \postcode{37996}, \state{TN}, \country{USA}}}

\abstract{Templated porous materials, such as MCM-41, due to the uniformity of their one-dimensional structure and scalability in synthesis, have emerged as an attractive medium for studying one-dimensional quantum fluids. However, the experimental challenge of synthesizing these materials with pore radii smaller than 15 $\dot{A}$ hinders the realization of a one-dimensional quantum liquid of helium within such systems, as the coherence length of helium is shorter than the pore radius. Recently, DelMaestro et. al. have preplated MCM-41 with Ar resulting in a reduction of the pore size and a softening of the adsorption potential allowing them to observe 1D Tomanga-Luttinger liquid like behavior. In this paper we present a novel method to obtain an even more ideal environment for studying the behavior of 1D $^4$He. We propose preplating MCM-41 pores with cesium (Cs) metal. The non-wetting nature of helium on a Cs-coated surface, coupled with the large atomic radius of cesium, creates an optimal environment for confining a quantum liquid of helium in one-dimensional geometry. We present preliminary measurements of  adsorption isotherms and Small Angle X-ray Scattering studies that reveal a reduction in pore radius upon preplating MCM-41 with Cs, demonstrating promising prospects for facilitating the realization of one-dimensional quantum fluids in templated porous materials.
}


\maketitle

%================================================

\section{Introduction}\label{sec1}
    
The impact of reduced dimensionality in quantum systems is a fundamental challenge in condensed matter physics, often leading to exotic phases like Dirac fermions, fractional quantum hall effect, and quantum spin liquids \cite{imambekov2012one}. The spatial dimension of a quantum system can be systematically reduced by imposing confinement on a scale smaller than the coherence length of the wavefunction. In systems like carbon nanotubes and electronic quantum wires, one-dimensional (1D) phenomena have been investigated by confining electrons to a length scales smaller than the Fermi wave vector using experimental techniques like electron-beam lithography\cite{bockrath1999luttinger,yao1999carbon,ishii2003direct,auslaender2005spin,jompol2009probing,laroche20141d,blumenstein2011atomically}.  Similarly, laser trap provides a confinement environment for atoms on the scale of thermal de-Broglie wavelength in ultracold atomic systems \cite{monien1998trapped,greiner2001exploring,paredes2004tonks,kinoshita2005local,haller2010pinning,sanchez2016mott,yang2017quantum}. In the case of superfluid helium, although the helium isotopes $^{3}$He and $^{4}$He have long been prototypes for precision tests of strongly interacting quantum matter and phase transition of bosons and fermions in two- and three-dimensions, the angstrom-scale coherence length $(\xi(T)\approx 1\textrm{nm})$ below the superfluid transition temperature $(T<T_{\lambda}\simeq 2.12\textrm{K})$ poses a challenge for engineering tunable physical systems capable of confining them in 1D geometry. \\

\noindent
Contemporary methods for physically confining superfluid helium can be broadly categorized into two techniques: nanofabrication and chemical synthesis. Nanofabrication involves the use of electron beams to etch short cylindrical pores with length less than 50 nm and radii ranging from 3 to 100 nm \cite{savard2009flow,savard2011hydrodynamics,duc2015critical}. Alternatively, heavy-ion bombardment of polymer films can be utilized to produce longer channels, with length between 1 to 100$\mu$m and radii from 15 to 200 nm \cite{velasco2012pressure,velasco2014flow,botimer2016pressure}. In both scenarios, the hydrodynamic behavior of the confined superfluid exhibits deviations from the typical three-dimensional pressure-driven flow observed in bulk systems, suggesting a transition towards one-dimensional confinement. The second approach, chemical synthesis, involves synthesizing silicates such as MCM-41 (Mobil composition of matter No. 41)\cite{kresge1992ordered}, which feature regular networks of hexagonal pores. The pore radii in these materials are determined by the specific reaction route and it is not continuously tunable, with the smallest possible radii being on the order of 1.5 nm. Quantum Monte Carlo simulations \cite{del2011he,del2012luttinger,kulchytskyy2013local,markic2015superfluidity} suggest that achieving truly 1D behavior of  $^{4}$He may require subnanometer radii, highlighting the need for systematic methods to further reduce nanopore sizes.\\

\noindent
Recently, Nichols et al.\cite{nichols2020dimensional} proposed preplating the pores of MCM-41 with argon to effectively reduce the pore diameter. Numerical simulations revealed a twofold effect of the argon monolayer. Firstly, it decreased the pore diameter from 15 Å to approximately 10 Å. Secondly, it lowered the adsorption potential at the pore surface from -170 K to -70 K. Simulations of $^4$He in these preplated pores indicated the formation of two solid layers, followed by a layer that could either be a dense liquid or solid. The final layer of helium was located at the center of the pore, with its density varying based on the amount of $^4$He present. These findings suggested that at certain fillings, the core liquid at the pore's center could achieve a density conducive to observing Tomonaga-Luttinger liquid (TLL) behavior. Del Maestro et al.\cite{del2022experimental} employed neutron scattering to investigate the static and dynamic structure of $^4$He in argon-preplated MCM-41. By combining these experimental measurements with quantum Monte Carlo simulations, they successfully demonstrated the emergence of one-dimensional TLL behavior.\\

\noindent
In this study, we present a nanoengineering approach to preplate MCM-41 with a single adsorbed layer of cesium, effectively reducing the pore radius to the subnanometer scale. Previous experiments \cite{nichols2020dimensional,del2022experimental} involving confinement of helium inside argon-preplated MCM-41 have demonstrated some 1D behaviors of helium. However, since the pore radius in those cases exceeded the coherence length of helium, complete 1D behavior was not observed. Here, we propose that the non-wetting properties of helium on cesium-coated surfaces, combined with the large atomic radius of cesium compared to argon, make this platform more suitable for achieving full 1D behavior. Preliminary experimental results from $\textrm{N}_{2}$ and $^4\textrm{He}$ adsorption isotherms confirm that cesium preplating effectively reduces the surface area and pore radius of MCM-41. Moreover, Brunauer-Emmett-Teller(BET) analysis suggests that helium does not wet the cesium-preplated MCM-41 surface.\\

%=================================================

\section{Background}\label{sec2}
MCM-41 is a mesoporous silicate molecular sieve, synthesized from close-packed, silica-coated micelles of a surfactant template. Its key characteristics include well-defined hexagonal pores, a narrow pore size distribution, minimal pore networking or pore blocking effects, and a high degree of pore ordering over micrometers length scale. Additionally, MCM-41 exhibits remarkable thermal, hydrothermal, chemical, and mechanical stability \cite{selvam2001recent}. The behavior of helium atoms inside such material is governed by the confinement potential. The shape and depth of the confinement potential strongly influence the energy levels, density distributions, and interaction dynamics \cite{nichols2020dimensional}.\\

\noindent
Helium, due to the weak He-He interaction potential and the large zero-point motion, doesn't form a liquid phase until 4 K, and substantial pressure is required to form a solid.  Due to these weak interactions $^4$He was viewed as the universal wetting agent since $^4$He is more strongly attracted to nearly anything else than to itself.  However, Cheng and Cole\cite{Cheng:1993oe} first suggested that this might not be a universal behavior. They pointed out that for certain rare-earth atoms, such as Cesium, the repulsive core potential extended to a large enough distance that a deep attractive potential well could not develop.  They noted that in this situation, $^4$He would be more strongly attracted to itself and would not wet a Cs coated surface.  This non-wetting behavior has subsequently been confirmed in numerous experiments. 

\begin{figure}[h]
\centering
\includegraphics[width=1.0\textwidth]{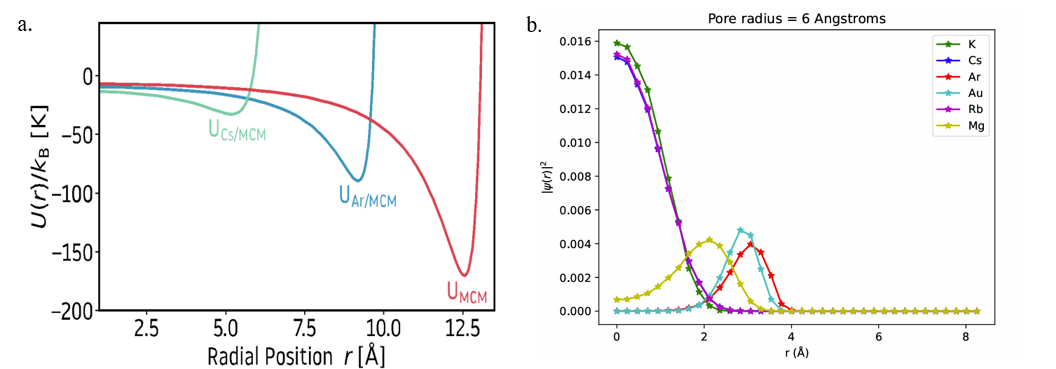}
\caption{a. Comparison of the confinement potential between MCM-41, argon-preplated MCM-41, and Cs-preplated MCM-41. b. Radial wave function from the center of the pore computed for 2 $^4\textrm{He}$ atoms inside MCM-41 pore preplated with different elements}\label{Confinement_potential_and_RDF}
\end{figure}

\noindent
Fig.{\ref{Confinement_potential_and_RDF}}a. represents a comparison of the confinement potential calculated using atomistic theory for MCM-41, Ar-preplated MCM-41, and Cs-preplated MCM-41. The confinement potential for MCM-41 exhibits a deep minima at approximately 12.5 \AA\ from the center of a pore, with a depth of $\sim$ -170K. Preplating MCM-41 with Ar results in a adsorption potential with a mimum at 8 \AA, with a potential depth of $\sim$~-70K. The numerical simulations of Nichols et. al. indicate that this minimum is still deep enough that solid layers form at the surface of the pores.\cite{nichols2020dimensional}\\

\noindent
Numerical simulations for Cs-preplated MCM-41 show the minima is located at approximately 5.0 \AA, with a depth of $\sim$ -30K. This indicates that preplating MCM-41 with Cs reduces the effective pore size to the coherence length scale of helium, thereby creating an ideal environment for confining helium in a 1D geometry. Additionally, the shallow potential well depth associated with Cs preplating indicates minimal interaction between helium and the Cs-preplated pore walls, resulting in an isolated environment conducive to 1D confinement of helium.\\

\noindent
Fig. {\ref{Confinement_potential_and_RDF}}b. depicts the simulated two-body radial wavefunctions for $^4\textrm{He}$ confined within MCM-41 pore preplated with various elements including K, Cs, Ar, Au, Rb, and Mg. The pore is modeled as a cylindrical structure made up of K, Cs, Ar, Au, Rb, and Mg with radius 6 \AA\ and a length of 25 \AA. For Cs, Rb, and K, the radial wave functions exhibit maxima near the center of the pore, decaying to zero beyond 2.5 \AA. This indicates that, for MCM-41 preplated with these elements, $^4$He atoms preferentially occupy the center region of the pore rather than near the wall, offering a more tunable environment for studying the 1D quantum liquid behavior of $^4$He.

%=================================================

\section{Results}\label{sec3}

%=================================================

\subsection{Preplating MCM41 with cesium}

Preplating Cs into the MCM-41 pores poses significant technical and experimental challenges. Cesium reacts explosively with water which presents a safety hazard and thus requires handling within an inert atmosphere. Additionaly, Cs is solid/liquid at room temperature with near zero vapor pressure, which means evaporation of a Cs into the MCM-41 pores must be done at relatively high temperatures. The temperature gradients must be carefully controlled to prevent completely filling the pore with Cs, desorption of the Cs film, or destabilization of the MCM-41 matrix. To overcome this challenge, we utilized a two-zone furnace which allowed us to independently control the temperature of the Cs reservoir and the MCM-41 substrate. Since the vapor pressure of Cs is around 1.0 Torr at 500K, we simultaneously raised the temperature of the source and sample to 550K, and slowly cool them down to room temperature.  

%=================================================

\subsection{Sample Characterization}

We characterized the Cs-preplated MCM-41 using Small Angle X-ray Scattering (SAXS) and adsorption isotherms.  The SAXS provides information on the spacing of the pores and the global distribution of the material in the pores.  Adsorption isotherms using helium and nitrogen provide information on the pore size and the adsorption potential.\\

\begin{figure}[h]
\centering
\includegraphics[width=0.8\textwidth]{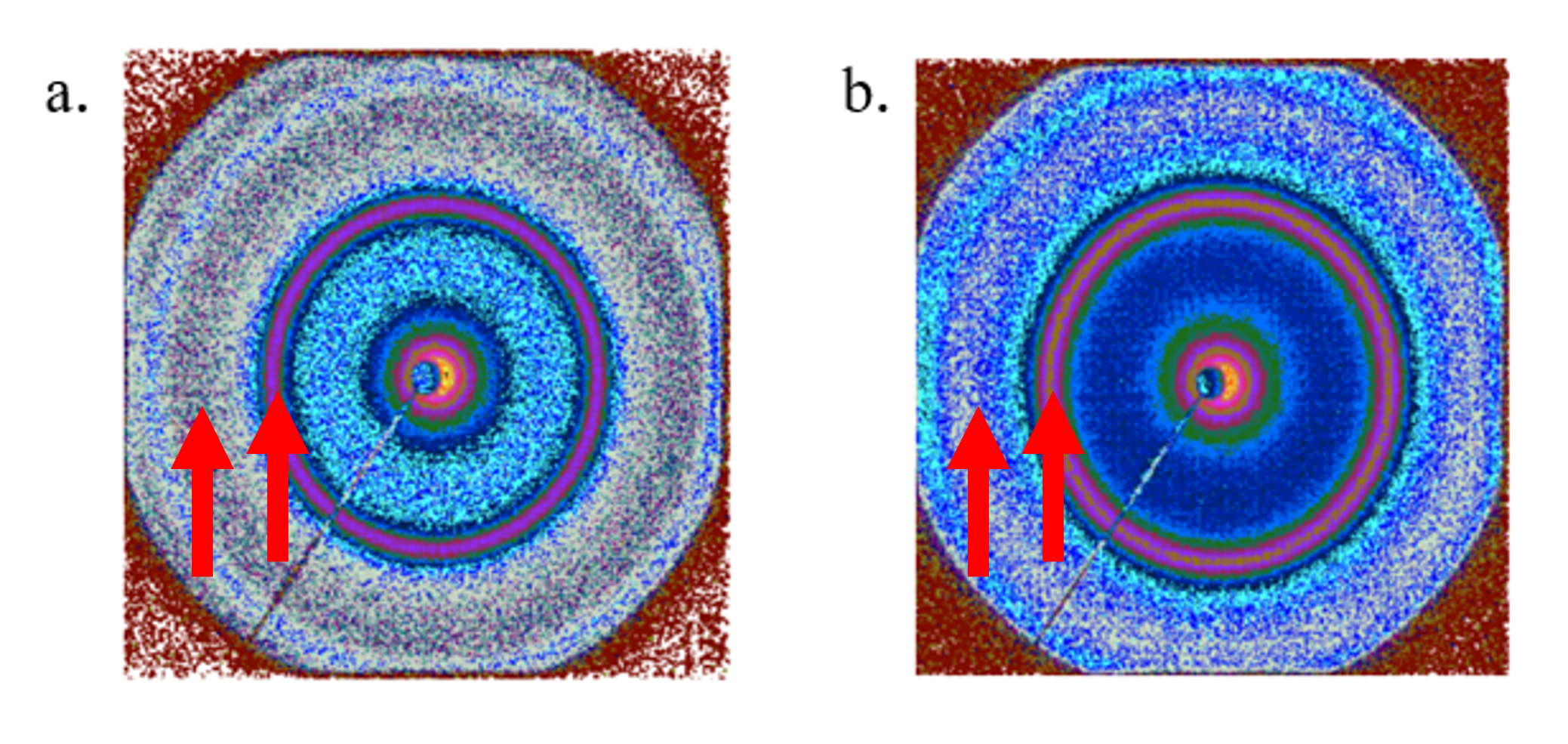}
\caption{Small angle X-ray diffraction pattern : a. MCM-41, and b. Cs-preplated MCM-41. The arrows mark the 10 and 11/20 diffraction peaks (the 11 and 20 are overlapping). The diffraction rings observed in both MCM-41 and Cs-preplated MCM-41 indicate that the 1D hexagonal structure of MCM-41 is also preserved in Cs-preplated MCM-41 }
\label{xray_diffraction}
\end{figure}

\noindent
SAXS measurements of bare and Cs coated MCM-41 are shown in Fig.\ref{xray_diffraction}. The SAXS pattern for both samples show two diffraction rings which are marked by the red arrows.  These rings occur at identical $Q$ values for both the unplated and plated samples.  This indicates that the overall hexagonal pore structure of the MCM-41 has not changed during the plating process.  However, the intensity of the diffraction rings differs between the two samples.  In particular the ratio of the intensities of the first and second rings is determined by the distribution of material in the sample.  Fig. \ref{xray_diffraction}a represents the structure of the as-synthesized MCM-41.  This is a matrix of SiO$_2$ with a hexagonal array of empty 1D pores.  The change in relative intensities of the diffraction peaks indicate shown in Fig. \ref{xray_diffraction}b is due to the Cs that is now present on the surface of the MCM-41 pore walls.\\

\begin{figure}[h]
\centering
\includegraphics[width=0.5\textwidth]{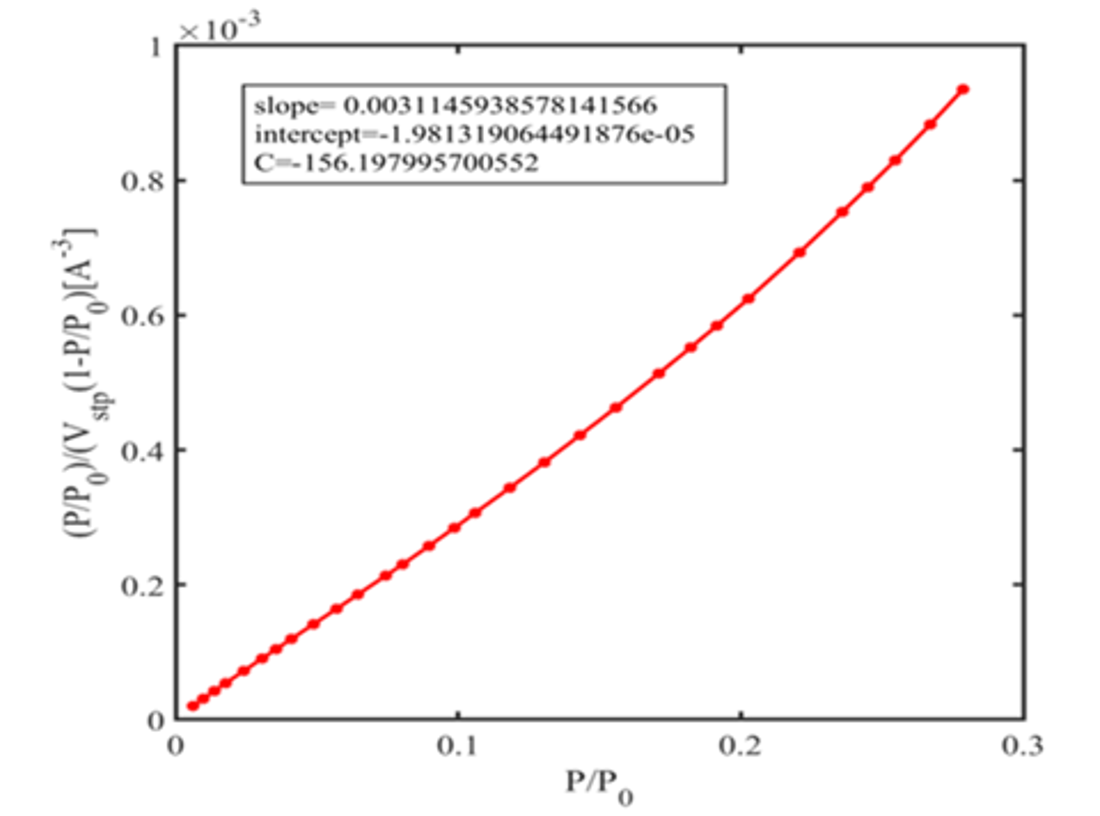}
\caption{BET analysis plot. The negative y-intercept of the BET isotherm suggests that helium does not wet cesium coated surface.}
\label{Isotherm_measurement}
\end{figure}

\noindent
Nitrogen adsorption isotherms were carried out at 77 K on both pristine MCM-41 and Cs-preplated MCM-41 to determine pore surface area and pore size. A Brunauer-Emmett-Teller (BET) analysis \cite{brunauer1938adsorption} of the N$_2$ isotherm gave a surface area of 965 {$m^2/g$} before preplating and 915 {$m^2/g$} after preplating.  The pore diameter size distribution was calculated using the Kruk-Jaroniec-Sayari method \cite{Jaroniec:1999mi} and was found to be Gaussian with a mean value of 37 \AA\ before plating and 30 \AA\ after plating.  The full-width at half-maximum was approximately 3 \AA\ in both cases. Thus, both pore surface area and radius are decreased with Cs adsorption. \\

\noindent
To study the physical adsorption of helium on Cs-preplated MCM-41, we measured helium isotherms at a temperature 5K. We then did Brunauer-Emmett-Teller (BET) analysis \cite{brunauer1938adsorption} to understand the interaction between helium and cesium. In general, the BET theory is not appropriate for helium isotherms due to the mobility of the liquid after the first few adsorbed layers.  However, it can be used for the first few layers to extract surface areas\cite{Band:1951}. The BET equation is given by:

\begin{equation}
    \frac{p/p_{0}}{v_{STP}[1-p/p_{0}]}=\frac{c-1}{v_{m}c}\left(\frac{p}{p_{0}}\right)+\frac{1}{v_{m}c}
    \label{BET_equation}
\end{equation}\\

\noindent
where, $p$, $p_{0}$ are the equilibrium and the saturation pressure of adsorbates respectively, $v_{STP}$ is the volume of adsorbed gas at STP, $v_{m}$ is the volume of the monolayer and $c$ is the BET constant. The BET constant is given by $c=\exp{\Delta E/RT}$, where $\Delta E$ is the difference between adsorbate-adsorbate and adsorbate-surface binding energy. The BET plot for helium adsorption is shown in Fig.\ref{Isotherm_measurement}. The most interesting feature is that c=-156 K -- a negative value.  This indicates that the BET theory breaks down for Cs plated MCM-41 implying that the $^4$He atoms are not wetting the Cs surface.

%=================================================

\section{Conclusion}
In this paper, we report a nanoengineering technique for preplating MCM-41 with cesium (Cs). The samples were characterized using X-ray diffraction (XRD) and nitrogen adsorption isotherm measurements. XRD analysis confirmed that the 1D hexagonal structure of MCM-41 is preserved after preplating with Cs. Nitrogen adsorption isotherms revealed a reduction in pore size for Cs-preplated  MCM-41 compared to unmodified MCM-41. Furthermore, BET analysis using helium adsorption isotherms indicated that Cs repels helium atoms toward the center of the pore. These findings suggest that Cs-preplated MCM-41 may serve as an ideal platform for studying the 1D behavior of superfluid helium. Ongoing experimental and theoretical investigations aim to further elucidate the properties of this system.

%=================================================

\bmhead{Acknowledgements}

This research was supported by the Department of Energy (DOE) under BES Award WS01054108 and the National Science Foundation (NSF) under Awards No. DMR-1809027 and No. DMR-1808440.

%=================================================

\bibliography{sn-bibliography}

\end{document}